\documentclass[10pt,journal]{IEEEtran} 
\usepackage{amsmath,amsfonts, amssymb, amsthm}

\newtheorem{theorem}{Theorem}
\newtheorem{corollary}{Corollary}

\usepackage{algorithmic}
\usepackage{algorithm}
\usepackage{bm, bbm}
\usepackage{array}
\usepackage[caption=false,font=normalsize,labelfont=sf,textfont=sf]{subfig}
\usepackage{textcomp}
\usepackage{stfloats}
\usepackage{url}
\usepackage{verbatim}
\usepackage{graphicx}
\usepackage{cite}
\usepackage{siunitx}
\usepackage{booktabs}
\usepackage{standalone}

\DeclareSIUnit{\dBSPL}{\dB_{\text{SPL}}}

\usepackage[svgnames]{xcolor} 
\definecolor{bluecite}{HTML}{0875b7}
\definecolor{RTX_neutral_medium}{RGB}{237,229,220}
\definecolor{RTX_neutral_light}{RGB}{237,236,231}
\definecolor{RTX_light_green}{RGB}{168,189,157}
\usepackage[unicode=true,
	bookmarksopen={true},
	pdffitwindow=true, 
	colorlinks=true, 
	linkcolor=bluecite, 
	citecolor=bluecite, 
	urlcolor=bluecite, 
	hyperfootnotes=false, 
	pdfstartview={FitH},
	pdfpagemode= UseNone]{hyperref}
\usepackage{orcidlink}
\usepackage[utf8]{inputenc}
\usepackage{pgfplots}
\DeclareUnicodeCharacter{2212}{−}
\usepgfplotslibrary{groupplots,dateplot}
\usetikzlibrary{patterns,shapes,arrows,calc,fit, decorations.pathreplacing}
\usepackage{tikzscale}
\pgfplotsset{compat=newest}
\begin{document}
\bstctlcite{IEEEexample:BSTcontrol}

\title{Minimum Processing Near-end Listening Enhancement}

\newcommand{\orcidAndreas}{\orcidlink{0000-0002-4199-5222}}
\newcommand{\orcidJan}{\orcidlink{0000-0002-3724-6114}}
\newcommand{\orcidLars}{\orcidlink{0000-0002-5145-8340}}
\newcommand{\orcidZheng}{\orcidlink{0000-0001-6856-8928}}
\newcommand{\orcidJesper}{\orcidlink{0000-0003-1478-622X}}

\author{Andreas~Jonas~Fuglsig\,\orcidAndreas,~\IEEEmembership{Student~Member,~IEEE,}
 Jesper~Jensen\,\orcidJesper,
 Zheng-Hua~Tan\,\orcidZheng,~\IEEEmembership{Senior~Member,~IEEE,}
 Lars~Søndergaard~Bertelsen\,\orcidLars,
 Jens~Christian~Lindof,~\IEEEmembership{Member,~IEEE,}
 Jan~Østergaard\,\orcidJan,~\IEEEmembership{Senior~Member,~IEEE}
\thanks{This work is partly funded by Innovation Fund Denmark Case no. 9065-00204B.}
\thanks{A. J. Fuglsig (email: ajf@\{rtx, es.aau\}.dk) is with RTX A/S, Denmark and the Department of Electronic Systems, Aalborg University, Denmark. J. Østergaard, J. Jensen and Z. Tan (email: \{jo, jje, zt\}@es.aau.dk) are with the Department of Electronic Systems, Aalborg University, Denmark. L. S. Bertelsen and J. C. Lindof (email: \{lsb, jcl\}@rtx.dk) are with RTX A/S, Denmark.}%
}

\markboth{Accepted by IEEE/ACM Transactions on Audio, Speech, and Language Processing}
{Fuglsig \MakeLowercase{\textit{et al.}}: Minimum Processing Near-end Listening Enhancement}


\maketitle

\begin{abstract}
The intelligibility and quality of speech from a mobile phone or public announcement system are often affected by background noise in the listening environment. By pre-processing the speech signal it is possible to improve the speech intelligibility and quality --- this is known as near-end listening enhancement (NLE). Although, existing NLE techniques are able to greatly increase intelligibility in harsh noise environments, in favorable noise conditions the intelligibility of speech reaches a ceiling where it cannot be further enhanced. Actually, the focus of existing methods solely on improving the intelligibility causes unnecessary processing of the speech signal and leads to speech distortions and quality degradations. In this paper, we provide a new rationale for NLE, where the target speech is minimally processed in terms of a processing penalty, provided that a certain performance constraint, e.g., intelligibility, is satisfied.
We present a closed-form solution for the case where the performance criterion is an intelligibility estimator based on the approximated speech intelligibility index and the processing penalty is the mean-square error between the processed and the clean speech. This produces an NLE method that 
adapts to changing noise conditions via a simple gain rule by limiting the processing to the minimum necessary to achieve a desired intelligibility, while at the same time focusing on quality in favorable noise situations by minimizing the amount of speech distortions. Through simulation studies, we show the proposed method attains speech quality on par or better than existing methods in both objective measurements and subjective listening tests, whilst still sustaining objective speech intelligibility performance on par with existing methods.
\end{abstract}


\IEEEpubidadjcol  

\section{Introduction}

\begin{figure}[!t]
  \centering
  \includegraphics[width=\columnwidth]{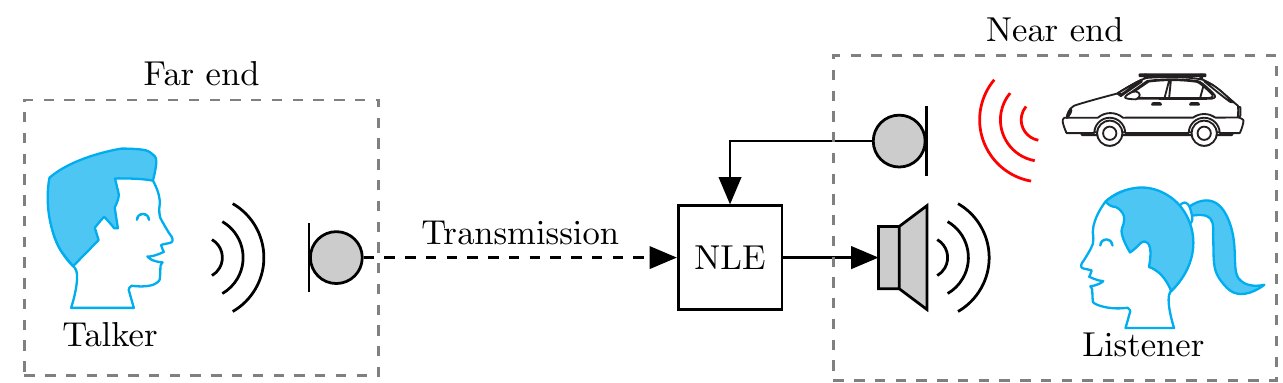}
  \caption{Basic principle of near-end listening enhancement (NLE), where the received far-end signal is processed prior to playout in a noisy environment.}
  \label{fig:nle_concept}
\end{figure}

Real-life speech communication, e.g., with mobile phones or public announcements, takes place in a large variety of often noisy places. Here, environmental noises such as cars, trains, construction work and other people talking may interfere with speech perception and degrade both the speech intelligibility (SI) and the speech quality (SQ).
In near-end listening scenarios, the noise sources are physically present in the environment where the listener is located, cf. the right-hand part of Fig.~\ref{fig:nle_concept}. Therefore, in the near-end scenario, we cannot extract the clean speech signal by removing noise from a noisy input signal as done in far-end speech enhancement, when there is noise in the left-hand side of Fig.~\ref{fig:nle_concept}, e.g., \cite{loizou_speech_2013,gannot_consolidated_2017}.
Instead, several other techniques exist for increasing the SI and SQ in noise by modifying the speech signal received from the far-end prior to playback in the noisy environment, and are known as Near-end Listening Enhancement (NLE).

Both SI and SQ are important factors for the listening experience, but the importance of SI and SQ changes depending on the acoustic situation~\cite{rennies_evaluation_2018,pricken_quality_2017, tang_study_2018}. Sometimes the requirements to SI and SQ may even be at a conflict and a trade-off must be made~\cite{tang_study_2018}. 
Increased intelligibility can lead to an increase in experienced SQ in noisy conditions~\cite{pricken_quality_2017,tang_study_2018}. That is, when listening in noise, the SI is an important contributing factor to the experienced SQ~\cite{tang_study_2018}. Even though NLE processing may introduce distortion to the target speech signal, which could lead to decreases in SQ, these speech distortions may be masked by the harsher environmental noise in these noisy situations~\cite{tang_study_2018, taal_speech_2014}. 
However, speech becomes naturally more intelligible as noise conditions improve~\cite{rennies_evaluation_2018,pricken_quality_2017, tang_study_2018}. Hence, in favorable noise conditions, the intelligibility of unprocessed speech approaches $\SI{100}{\percent}$ so that it cannot be further enhanced~\cite{rennies_evaluation_2018}. In fact, further excessive or unnecessary processing of the speech signals may lead to speech distortions and quality degradations because the noise can no longer mask these processing distortions~\cite{tang_study_2018}. Therefore, it may be useful to adapt the NLE processing to the noise situations, such that it can be activated and deactivated in the best possible way~\cite{tang_study_2018}.

Traditionally, the goal of NLE algorithms is solely to maximize the SI for the listener in the noisy environment by modifying the time-frequency characteristics of the input speech such that it is not masked by the environmental noise, cf. the reviews in \cite{cooke_evaluating_2013,cooke_listening_2014,kleijn_optimizing_2015} and the contributions to the Hurricane challenges~\cite{cooke_intelligibility-enhancing_2013, rennies_intelligibility-enhancing_2020}.
%
Existing NLE techniques can be categorized into two main classes. 
The first one is the heuristic or expert driven approaches, e.g., 
\cite{niederjohn_enhancement_1976,hall_intelligibility_2010,nathwani_speech_2021,tantibundhit_new_2007,
jokinen_intelligibility_2017, zhang_spectral_2018, li_mapping_2020,
zorila_speech--noise_2012, chermaz_sound_2020,
sauert_near_2006-1, sauert_near_2006,
niermann_near-end_2016, niermann_listening_2021
}. These consider various approaches regarding, e.g., formants, transients, Lombard speech, spectral shaping and dynamic range compression, or plain audibility. Although the class of heuristically based NLE techniques provide significant improvements in SI, they are often not derived according to specific objective optimality criteria. Instead, they are non-parametric and based on subjective expert experiences and knowledge. Hence, they cannot claim optimality and maybe are not adapting well to changing environments.

The second class of NLE algorithms is based on the main idea of manipulating the input speech such that a target intelligibility metric is maximized when the noise conditions are known. 
One of the most widely used optimization targets for SI enhancement is the Speech Intelligibility Index (SII)~\cite{american_national_standards_institute_methods_2017} or variations thereof, which have been used for NLE in numerous studies, e.g., \cite{sauert_recursive_2010, taal_optimal_2013,schepker_speech--noise_2015, hendriks_optimal_2015, niermann_joint_2017,sauert_near-end_2014, fuglsig_joint_2022}. The SII based approaches \cite{sauert_recursive_2010,taal_optimal_2013} show good performance but fall behind the state-of-the-art heuristic~\cite{zorila_speech--noise_2012} in subjective tests~\cite{cooke_intelligibility-enhancing_2013, rennies_intelligibility-enhancing_2020}, because the optimization targets do not correlate well with subjective intelligibility across varying noises and degradations~\cite{van_kuyk_evaluation_2018}. Furthermore, some methods, e.g., \cite{sauert_recursive_2010,taal_optimal_2013}, require solving an optimization problem in real-time with varying execution time~\cite{niermann_near-end_2016}. The SII based solutions also rely on simplifying assumptions about frequency gains being constant across frequency subbands~\cite{fuglsig_joint_2022}. Recently, deep neural network (DNN) based approaches \cite{li_imetricgan_2020,li_multi-metric_2021} have been able to optimize more advanced measures such as the extended short-time objective intelligibility
(ESTOI) measure~\cite{jensen_algorithm_2016} that correlate more with subjective tests than the simpler metrics such as SII~\cite{van_kuyk_evaluation_2018}. However, DNN methods come at the cost of significant memory usage, black-box solutions, and possible problems in generalizing to new acoustic scenarios.

We note for near-end listening with headphones, adaptive noise cancellation (ANC) techniques\cite{kuo_active_1999,george_advances_2013} may be employed, which --- instead of processing the target speech signal --- aim at adaptively cancelling the noise by adding an anti-phase noise component to the speech signal before playout in the noisy environment~\cite{li_near-end_2019, cheng_speech_2018}. However, ANC with classic adaptive filtering has generally been insufficient for improving intelligibility outside headphone use until the use of DNNs~\cite{li_near-end_2019}. Hence, NLE based on speech modification is still the predominant approach.

Common for the objective SI metrics such as SII and the Glimpse model~\cite{cooke_glimpsing_2006} is that audibility is the decisive factor of intelligibility. Therefore, the principle for all NLE algorithms is to adjust the SNR for the perceptually important parts of the speech~\cite{li_near-end_2019}. Hence, the simplest solution to the NLE problem is to increase the power of the clean speech signal until the noise is sufficiently masked, i.e., increasing the SNR (almost indefinitely).
However, such an increase in speech power may lead to various problems, most significantly possible hearing damage to the near-end listener, but also problems with loudspeaker overload and unpleasant playback levels~\cite{schepker_speech--noise_2015}. Therefore, many NLE approaches take a power constraint into account, such that an equal power constraint is maintained between the unprocessed and processed clean speech signal. However, this also considerably limits the potential for increasing the SI~\cite{sauert_near_2009,niermann_listening_2021}.

Existing methods are designed to always achieve the desired output power level. Thus, the power is always increased (or decreased) to the maximum allowed level even though this may not be necessary and may lead to excessive processing and a decrease in SQ or SI~\cite{cooke_listening_2014, tang_study_2018}. Thus, increasing the allowed output power too much, in order to overpower the noise, may cause problems for SQ. 
Hence, we need to control the amount of processing in a different way, such that we can achieve both good SI and SQ depending on the noise situation.

In this work, we take inspiration from the area of far-end noise reduction. Particularly, the work of Zahedi et al.~\cite{zahedi_minimum_2021}, where the concept of \emph{minimum processing beamforming} is proposed. In this new rationale, the goal is to ensure a minimum level of SI performance, while guiding the noise reduction capabilities of the beamforming towards a particular performance; ambient-preserving, aggressive noise reduction or standard Wiener filtering.

In this paper, we propose the concept of \emph{minimum processing near-end listening enhancement}. The proposed concept provides a more general formulation of the NLE problem that focuses on both SI and SQ in an adaptive manner.
The concept is based on minimizing a speech processing penalty subject to a certain intelligibility performance constraint.
The point is no longer only to maximize SI but instead process the target speech signal just enough to achieve a minimum desired intelligibility in the given noise conditions, while minimizing the amount of speech distortions in favorable noise situations. 
Thus, the concept is adapting to the noise conditions by increasing SI when needed and ensuring an automatic focus on SQ, when the SI is already sufficient by automatically reducing the amount of processing when the desired SI is achieved.

We present an exemplary case study where the processing penalty is the mean-square error (MSE) between the processed near-end signal and the clean speech, and the performance criterion used in the optimization is an intelligibility estimator based on the approximated SII (ASII)~\cite{taal_optimal_2013}. While it is common in SII based works to assume that gains in the same subband are the same, we do not assume this upfront in this work. Instead, the optimization problem in this work is specifically built as a function of the gains in STFT domain to formulate a more general optimization problem without this common assumption. The case study shows we are able to adapt to changing noise conditions via a simple gain rule that is computationally efficient and which does not require online optimization. Furthermore, it is a result of our work that for the given case study the gains within a subband are the same, thus
the derived solution mathematically validates the simplifying assumption of constant gains made in existing SII based works.

Finally, the proposed NLE method achieves SQ performance on par or better than existing methods in both objective measurements and subjective listening tests, while maintaining objective SI performance on par with existing methods for a wide range of SNRs.

The rest of the paper is organized as follows. In Section~\ref{sec:signal_model} we introduce our signal model. Section~\ref{sec:minproc_NLE} introduces the proposed minimum processing NLE and we make our case study. Section~\ref{sec:practical_considerations} introduces practical considerations and summarizes the algorithm. Section~\ref{sec:objectiv_perform} presents the experimental setup and objective SI and SQ performance. Section~\ref{sec:subjec_qual_test} presents a subjective SQ listening test and discusses the results. We conclude the paper in Section~\ref{sec:conclusion}.




\section{Signal Model}\label{sec:signal_model}
\begin{figure}[!t]
  \centering
  \includegraphics[width=\columnwidth]{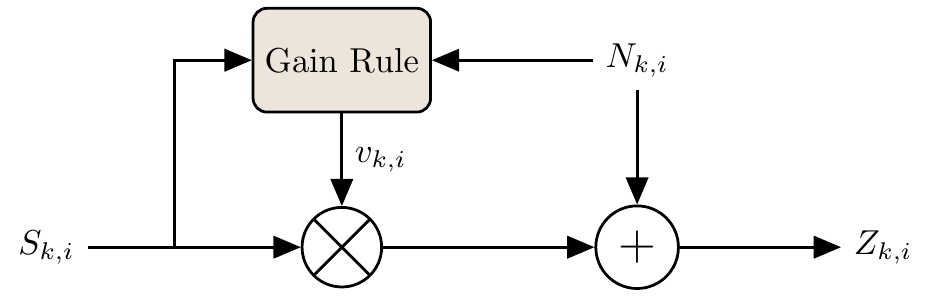}
  \caption{Block diagram representation of frequency domain NLE signal model.}
  \label{fig:nle_block}
\end{figure}

%
In this work we consider a signal model where speech and noise are represented in the complex short-time discrete Fourier transform (STFT) domain, cf. Fig.~\ref{fig:nle_block},
Here $S_{k,i}$ is the clean speech signal received from the far-end in frequency-bin $k$ and time-frame $i$. In most general settings the far-end signal is noisy, however, we assume an adequately clean version is available or can be achieved through proper noise reduction at the far-end. $N_{k,i}$, is the time-frequency representation of the additive near-end noise. As is common in the literature, we assume  the speech, $S$, and noise, $N$, are uncorrelated of each other. To increase the SI (and SQ) of the signal received by the near-end listener, the speech signal, $S$, is linearly pre-processed prior to play out in the near-end environment. Thus, we assume that the signal presented to the near-end listener in STFT domain, $Z_{k,i}$, follows
\begin{align}
  Z_{k,i} = v_{k,i}S_{k,i} + N_{k,i},
\end{align}
where $v_{k,i}$ denotes the linear pre-processing gains. The NLE gains constitute the core of NLE, and determining a proper NLE gain rule is the core task of NLE.

\subsection{Subband Model}
We focus on NLE based on (the perceptually driven) ASII for speech intelligibility prediction. ASII is part of a family of well-known speech intelligibility and quality predictors such as ESTOI~\cite{jensen_algorithm_2016}, STOI~\cite{taal_algorithm_2011}, SII~\cite{american_national_standards_institute_methods_2017}, extended SII (ESII)~\cite{rhebergen_speech_2005}, the hearing-aid speech perception index (HASPI)~\cite{kates_hearing-aid_2014}, the hearing-aid speech quality index (HASQI)~\cite{kates_hearing-aid_2014-1} and perceptual evaluation of speech quality (PESQ)~\cite{itu-t_recommendation_2001}, that all mimic aspects of human speech perception. That is, signals are analyzed in, e.g., octave bands, fractional octave bands or critical bands similar to the human ear. In general, these bands are referred to as subbands. Perceptually motivated subbands may be defined such that multiple frequency bins contribute to the same and/or multiple subbands, with an individual weight for each frequency bin-subband pair. We index subbands with index $j$ and frequencies with index $k$.

To illustrate the use of subbands in this work, let $\omega_{j,k}$ denote the non-negative filter weights that implement the $j$'th subband filter, $\mathbb{B}_j$ denote the set of frequency bins that contribute to the $j$'th subband with $j \in \{1,\ldots, J\}$ and $J$ the total number of subbands. Then,
the clean speech spectrum level within one subband, $j$, and time-frame, $i$, is defined as
\begin{equation}
  \sigma_{\mathcal{S}_{j,i}}^2\triangleq \sum_{k \in \mathbb{B}_j} \omega_{j,k} \sigma_{S_{k,i}}^2,
\end{equation}
where $\sigma_{S_{k,i}}^2$ is the clean speech spectrum level in time-frequency bin $k,i$. For the sake of brevity and ease of reading we assume any normalization of the subband filter weights, $\omega_{j,k}$, is already included in the weights. We elaborate further on the definition of $\omega_{j,k}$ and the relation between subbands and frequency bins in Appendix~\ref{app:subband_def}.

Finally, we assume the speech and noise are processes of complex random vectors comprised of the STFT coefficients. In practice one could estimate the statistics of these processes online, and the mathematical framework can be applied to the time-varying case on a per time-frame basis. Therefore, for brevity of notation we disregard the time-index, $i$, and assume we are considering a certain time frame $i$, unless otherwise is stated.


\section{Minimum Processing Near-End Listening Enhancement}\label{sec:minproc_NLE}
\subsection{Concept}

For a particular subband $j$, we denote by $\bm{S}_j \in \mathbb{C}^{\lvert \mathbb{B}_j\rvert}$ the vector containing all $S_k$ for $k \in \mathbb{B}_j$, where $\lvert \mathbb{B}_j\rvert$ denotes the number of frequency bins in the $j$'th subband. Similarly, we create the vector $\bm{Z}_j$ by stacking $Z_k$ for $k\in \mathbb{B}_j$. Furthermore, let $\mathcal{D}_j(\cdot, \cdot)$ and $\mathcal{I}_j(\cdot, \cdot)$ be finite non-negative functionals, where $\mathcal{D}_j\left(\bm{S}_j, \bm{Z}_j\right)$ measures the distortion (processing penalty) between the clean speech signal, $S$, and the signal presented to the near-end listener, $Z$, and $\mathcal{I}_j\left(\bm{S}_j, \bm{Z}_j\right)$ is an intelligibility or performance estimator for the NLE in subband $j$.
Then, the minimum processing near-end listening enhancer in subband $j$ is defined as the solution to the following optimization problem:
\begin{equation}
  \begin{array}{lll}
      \displaystyle \min_{\{v_k\} \in \mathbb{R}_+,~k \in\mathbb{B}_j} &\mathcal{D}_j\left(\bm{S}_j, \bm{Z}_j\right) & \mbox{s.t.}\quad \mathcal{I}_j\left(\bm{S}_j, \bm{Z}_j\right)\geq I'_j. 
  \end{array}\label{eq:gen_minproc_problem}
\end{equation}
The term $I'_j$ is defined as 
\begin{equation}
  I_j' = \min \left(I_j, I_j^{\mbox{max}}\right),
\end{equation}
where $I_j$ is a desired minimum requirement on the NLE intelligibility performance $\mathcal{I}_j\left(\bm{S}_j, \bm{Z}_j\right)$ and $I_j^{\mbox{max}}$ is the maximum achievable performance when disregarding the processing penalty $\mathcal{D}_j\left(\bm{S}_j, \bm{Z}_j\right)$, i.e., when the performance $\mathcal{I}_j\left(\bm{S}_j, \bm{Z}_j\right)$ is maximized in an unconstrained manner. We highlight the generality of the problem formulation in \eqref{eq:gen_minproc_problem}, in that $\mathcal{I}_j\left(\bm{S}_j, \bm{Z}_j\right)$ is a general ``performance criterion" which could reflect any type of performance aspect of interest to the application, e.g., SI, SQ or other measures of interest.

\subsection{Case Study}
In this paper, we study the case, where the processing penalty, $\mathcal{D}_j$, is the MSE criterion, and the performance criterion, $\mathcal{I}_j$, is an intelligibility estimator based on the ASII~\cite{taal_optimal_2013}. We solve the problem analytically for any given minimum performance constraint and subband definition. Furthermore, we show how we can select $I_j$ to achieve a desired total minimum performance across all subbands, $A^*$, i.e, we guarantee that the optimum solution satisfies

\begin{equation}
  \sum_{j}\mathcal{I}_j\left(\bm{S}_j, \bm{Z}_j\right) \geq A^*.
\end{equation}

\subsubsection{Processing Penalty}\label{sec:proc_penal}
To ensure the processing does not distort speech excessively and the playback volume is not increased infinitely, we introduce a processing penalty~\cite{zahedi_minimum_2021}. We consider a MSE processing penalty, where the MSE is evaluated in subband domain instead of DFT domain to ensure compatibility with the minimum processing NLE formulation in \eqref{eq:gen_minproc_problem}. That is, the $j$'th subband MSE processing penalty is
\begin{equation}
  \mathcal{D}_j(\bm{S}_j, \bm{Z}_j)=\sum_{k\in \mathbb{B}_j}\omega_{j,k}\left( 1 - v_{k} \right)^2 \sigma_{S_{k}}^2.\label{eq:mse_penal}
\end{equation}
See Appendix~\ref{app:mse_proc_penalty} for the derivation.

\subsubsection{Performance Criterion}\label{sec:perform_crit}


We consider, as an example of the proposed frame work, a performance criterion based on the ASII~\cite{taal_optimal_2013}. The original work on minimum processing in \cite{zahedi_minimum_2021} considered a performance criterion based on the SII~\cite{american_national_standards_institute_methods_2017}. Both SII and ASII consider intelligibility to be a weighted sum of intermediate measures of the audibility of speech in a subband as a function of the subband SNR. In SII, the band audibility is determined by a function where the long-term SNR is log-transformed and clipped between $\SI{-15}{\dB}$ and $\SI{+15}{\dB}$ and normalized to range between zero and one. In ASII, the audibility is determined by a non-linear approximation of the log-transform and clipping via a sigmoidal function of the subband SNR. The advantage of using the ASII is that its nonlinear approximation of the SII provides a nice mathematical tractability circumventing the need for dealing with clipping of SNRs.
Based on the ASII, have the following performance criterion for the $j$'th subband,
\begin{equation}
   \sum_{k\in \mathbb{B}_j} \omega_{j,k} v_{k}^2 \sigma_{S_{k}}^2  \geq \sigma_{\mathcal{N}_{j}}^2 I_j^\xi,\label{eq:SNR_crit}
\end{equation}
where $I_j^\xi \triangleq \frac{I_j}{1-I_j}$, and $I_j$ is the desired subband audibility performance.
The details are shown in Appendix~\ref{app:asii_perform_criterion}.


\subsection{Optimization Problem and Solution}\label{sec:opt_prob_sol}

Joining the results of \eqref{eq:mse_penal} and \eqref{eq:SNR_crit}, the minimum processing NLE problem \eqref{eq:gen_minproc_problem}, which we consider, and its solution is given in the following theorem.

\begin{theorem}\label{theo:minproc_solution}
The minimum processing NLE problem \eqref{eq:gen_minproc_problem} with MSE processing penalty \eqref{eq:mse_penal} and ASII performance criterion \eqref{eq:SNR_crit} is
    \begin{equation}
    \begin{array}{ll}
      \displaystyle \min_{\{v_k\} \in \mathbb{R}_+, k\in\mathbb{B}_j} & \sum_{k\in \mathbb{B}_j}\omega_{j,k}\left( 1 - v_{k}\right)^2 \sigma_{S_{k}}^2,\\
      \mbox{subject to} &  \sum_{k\in \mathbb{B}_j}\omega_{j,k} 
      v_{k}^2 \sigma_{S_{k}}^2  \geq \sigma_{\mathcal{N}_{j}}^2 I_j^\xi.
    \end{array}~\label{eq:NLE_minproc_problem}
  \end{equation}
The optimal  \emph{minimum processing NLE} gains are
  \begin{equation}
    v_{k,j}^{\text{MP}} = \begin{cases}
      1 & \text{if}~\sigma_{\mathcal{S}_{j}}^2  \geq \sigma_{\mathcal{N}_{j}}^2 I_j^\xi\\
      \sqrt{\frac{\sigma_{\mathcal{N}_{j}}^2 I_j^\xi}{\sigma_{\mathcal{S}_{j}}^2}}  & \text{otherwise}
    \end{cases}, \quad \forall k \in \mathbb{B}_j.\label{eq:optimum_gains}
  \end{equation}
\end{theorem}
In Appendix~\ref{app:opt_sol_deriv} we derive the \emph{minimum processing NLE}, i.e., the optimal gains in \eqref{eq:NLE_minproc_problem}.

We immediately have the following result regarding the optimal minimum processing NLE subband gains \eqref{eq:optimum_gains}.
  
\begin{corollary}
  The optimum gains for the minimum processing NLE \eqref{eq:NLE_minproc_problem} are equal for all frequencies within a subband $j$.   
\end{corollary}

This is an important result because most work on subband SNR based enhancement make this assumption for convenience in the joint optimization across all subbands~\cite{fuglsig_joint_2022, taal_optimal_2013, khademi_intelligibility_2017,niermann_joint_2017, kleijn_optimizing_2015}. However, as we show, it can be deduced from \eqref{eq:optimum_gains} that the optimal solution does not depend upon $k$, and it is optimal to have the same gain across the entire subband when optimizing for each subband individually.

For the optimal gain \eqref{eq:optimum_gains}, the first case takes effect when the unprocessed speech signal already satisfies the performance constraint, i.e., the speech has an intelligibility at or above the desired level. Hence, there is no reason to process the speech signal and the optimal solution is to do nothing, i.e., $v_{k,j}^{\text{MP}} = 1$ for $k \in \mathbb{B}_j$, and thus maximize SQ by minimizing the MSE.

In the second case, the unprocessed speech is below the desired audibility level or equivalently the desired SNR. The optimal gain then increases the speech power such that the subband SNR is exactly at the desired level. That is, the speech power in the $j$'th subband is increased just enough to satisfy the subband audibility constraint. This illustrates how the optimal solution provides the minimum required processing necessary to achieve the desired intelligibility.
Thus, the two cases of \eqref{eq:optimum_gains} illustrate how our approach has the advantage of achieving a desired intelligibility level, while automatically adjusting the processing to minimize processing artifacts.

Squaring the gains of \eqref{eq:optimum_gains}, $(v_{k,j}^{\text{MP}})^2$, we can see that the optimal processed speech power \eqref{eq:proc_subband_power} is increased proportionally to the near-end noise power. This is also the naturally expected solution to ASII and SII based NLE approaches given sufficient ability to increase speech power~\cite{niermann_listening_2021, taal_optimal_2013,sauert_recursive_2010}. Thus, we provide a mathematical justification for the heuristic approach of increasing speech power proportionally to the noise in~\cite{niermann_listening_2021}. 

Depending on the choice of subband definition, the subbands may overlap such that multiple frequencies contribute to multiple subbands indexed by $\mathbb{F}_k$. Therefore, the optimum gain, $v_{j,k}^*$ may also contribute to multiple subbands as indicated by the dependence on both $j$ and $k$. Using the weights of these contributions, $\omega_{j,k}$, we can weigh the filter gains through the subband filters. Thus, we have the following corollary.
\begin{corollary}
  The NLE processor in frequency bin $k$ is given as
  \begin{equation}
    v_{k}^{\text{MP}} = \sqrt{\sum_{j\in \mathbb{F}_k}\omega_{j,k}\left(v_{k,j}^{\text{MP}}\right)^2}.\label{eq:subband_filter_gains}
  \end{equation}
\end{corollary}
From this corollary, we see for overlapping subbands, that gains that were previously equal do not remain the same after being projected back to STFT domain.

\section{Practical Considerations}\label{sec:practical_considerations}
\subsection{Preventing excessive sound levels}
The proposed optimal minimum processing increases speech power to the necessary level to achieve the desired intelligibility. However, this may lead to speech levels that are unpleasant for the user. In extreme cases, if not controlled, such high signal levels could even cause hearing damage. Usually in most NLE work, this is prevented by a power equality constraint~\cite{taal_optimal_2013,niermann_listening_2021}. To prevent excessive output levels, in our work, we put an upper limit on the maximum processed speech power within each subband as described in~\cite[Sec. 2.2.5]{sauert_near-end_2014}. The idea is to limit the gain applied to each subband, $j$, such that the resulting subband power of the enhanced speech signal does not increase beyond a maximum subband power, $P^{\text{max}}_S$. That is,
\begin{equation}
  \overline{v}_{k,j}^{\text{MP}} = \operatorname{min}\left\{v_{k,j}^{\text{MP}}, v_{j,\text{max}}\right\} \quad \forall k \in \mathbb{B}_j,\label{eq:limit_gains}
\end{equation}
with maximum subband gain
\begin{equation}
  v_{j, \text{max}} = \sqrt{\frac{P^{\text{max}}_S}{\sigma_{\mathcal{S}_j}^2}}.\label{eq:max_gain}
\end{equation}
This is done prior to the subband filtering of the gains in \eqref{eq:subband_filter_gains}. Thus, $\overline{v}_{k,j}^{\text{MP}}$ is inserted on the right-hand side of \eqref{eq:subband_filter_gains}.

\subsection{Choosing the intelligibility limit per band}

The proposed minimum processing NLE as well as the work of \cite{zahedi_minimum_2021} considers a performance constraint, $I_j$, for each particular subband $j \in \{1,\ldots,J\}$. Thus, the processed speech achieves a desired performance in each particular subband. This target must be decided upon for each of the $J$ subbands before processing. In this section, we show how, when using the ASII criterion, we can select and achieve a single total target intelligibility, $A^*$, instead of having to select a desired performance constraint for each of the $J$ subbands. This single total target intelligibility is then converted to a per band intelligibility target.

Let $I_j$ be a given minimum requirement on the performance in a particular subband, $j$, i.e., we require 
\begin{equation}
  f(\xi_j) \geq I_j.
\end{equation}
Then the processed ASII score satisfies
\begin{align}
    \text{ASII} =  \sum_{j}\gamma_j f(\xi_j)\geq  \sum_{j}\gamma_j I_j. \label{eq:asii_lb}
\end{align}
That is, given that the minimum performance requirement is met for each band, the processed ASII is greater than or equal to a weighted sum of the performance criterions, where the weights are the band importance functions.

The band importance functions, $\gamma_j$, in the SII~\cite{american_national_standards_institute_methods_2017} describe how some subbands are more important for intelligibility than others. Therefore, it is natural to require higher audibility limits for the more important subbands and lower audibility limits for the less important bands. We can achieve this by using the band importance functions.

Let $A^*$ be the required minimum total ASII performance, and let
the subband audibility limits be given as
\begin{equation}
    I_j = \frac{A^* \gamma_j}{\sum_i \gamma_i^2}.\label{eq:weighted_I_lim}
\end{equation}
Then by inserting in \eqref{eq:asii_lb} the resulting ASII score satisfies
\begin{align}
    \text{ASII} &\geq  \sum_{j} \gamma_j \frac{A^* \gamma_j}{\sum_i \gamma_i^2}
    = \frac{A^*}{\sum_i \gamma_i^2}\sum_{j} \gamma_j^2
    = A^*,
\end{align}
that is, the total processed ASII score is lower bounded by the required performance. This means we can guarantee a total ASII greater than or equal to a target ASII, $A^*$.

As a special case, we see that removing the weights and selecting a fixed audibility limit across subbands ${I_j = A^*,~\forall j}$ also achieves  $\text{ASII} \geq A^*$. In our simulations we use the weighted audibility limits, as these have shown a slightly better performance than having the same fixed limit for each subband.



\subsection{Estimating statistics}
When processing time-varying signals, such as speech and non-stationary noise, the statistics of the signals change over time. Therefore, the statistics must be estimated and updated in time using, e.g., recursive averaging. However, if the statistics are updated too fast or abruptly, the optimal gains in \eqref{eq:optimum_gains} can change suddenly between time-frames, which causes audible distortions to the target speech signal. To circumvent this, it is common in the literature, e.g.,~\cite{niermann_listening_2021,zahedi_minimum_2021}, to use slowly time-varying processing, where the recursive averaging is across several frames (seconds). Therefore, in this work, we let the average energy per DFT bin and critical band be based on a long-term average over several short-time frames,
\begin{align}
    \sigma_{S_k}^2 &\triangleq \frac{1}{I}\sum_i  \lvert S_{k,i} \rvert^2, \label{eq:avg_power_time}\\
    \sigma_{\mathcal{S}_j}^2&\triangleq \sum_k\omega_{j,k}\sigma_{S_k}^2,\label{eq:filt_into_bands}
\end{align}
where $I$ is the total number of frames. Similar expressions hold for the near-end noise signal $N_{k,i}$. Thus, the estimated statistics in this work do not change over time, and the implemented processing is time-invariant. For time-varying estimates, \eqref{eq:avg_power_time} should be updated to a recursive or moving average.

\subsection{Algorithm summary}
\begin{figure}[!t]
  \centering
  \tikzset{%
  block/.style    = {draw, thick, rectangle, minimum height = 3em,
  minimum width = 3em, align=center, rounded corners,fill=RTX_neutral_medium},
  input/.style    = {coordinate}, 
  output/.style   = {coordinate}, 
}
\begin{tikzpicture}[auto, thick, node distance=2.1cm, >=triangle 45,scale=0.7, every node/.style={scale=0.7}]
  \draw
    node[align=center] at (0,0) (speech_pow) {Speech spectrum\\$\sigma_{S_{k,i}}^2$}
    node[right of=speech_pow,align=center, node distance=3cm] (noise_pow) {Noise spectrum\\$\sigma_{N_{k,i}}^2$}
    node[right of=noise_pow, align=center, node distance=3.5cm, yshift=1mm] (target_ASII) {Target intelligibility\\$A^*$}
    node [below of=speech_pow, block, node distance=2cm] (speech_filt) {Subband\\ filtering \eqref{eq:filt_into_bands}}
    node [below of=noise_pow, block, node distance=2cm] (noise_filt) {Subband\\ filtering \eqref{eq:filt_into_bands}}
    node[block] at (noise_filt -| target_ASII) 
    (target_weight) {Weight audibility\\ limit \eqref{eq:weighted_I_lim}}
    node[block, below of=target_weight] (SNR_lim) {SNR limit \eqref{eq:SNR_lim_def}}
    node[block]  at (noise_filt |- SNR_lim) (optimal_gain) {Optimal\\ gains \eqref{eq:optimum_gains}}
    node[block, below of=optimal_gain] (limit_gain) {Limit sound level \eqref{eq:limit_gains}}
    node[block, below of=limit_gain] (filter_gain) {Get frequency bin gains \eqref{eq:subband_filter_gains}}
    node[below of=filter_gain, node distance=2cm, align=center] (final_gain) {Minimum processing NLE gains\\$v_{k,i}^\text{MP}$}
    ;
    \draw[->](speech_pow) -- (speech_filt);
    \draw[->](noise_pow) -- (noise_filt);
    \draw[->](target_ASII) -- (target_weight);
    \draw[->](speech_filt) |- node[above, pos=.7] {$\sigma_{\mathcal{S}_j}^2$} (optimal_gain);
    \draw[->](noise_filt) -- node[right] {$\sigma_{\mathcal{N}_j}^2$} (optimal_gain);
    \draw[->](target_weight) -- node[right] {$I_j$} (SNR_lim);
    \draw[->](SNR_lim) -- node[above] {$I_j^\xi$} (optimal_gain);  
    \draw[->](optimal_gain) -- node[right] {$v_{k,j}^\text{MP}$} (limit_gain);  
    \draw[->](limit_gain) -- node[right] {$\overline{v}_{k,j}^\text{MP}$} (filter_gain);  
    \draw[->](filter_gain) -- (final_gain);  

    \node[color=gray,thick, dashed, draw, rectangle, xshift=.4cm, minimum width=9.4cm, minimum height=1.5cm, label={Input}] at (noise_pow) {};
    \node[color=gray,thick, dashed, draw, rectangle, minimum width=5cm, minimum height=1.2cm, label={below:{Output}}] at (final_gain) {};
  \end{tikzpicture}
  \caption{Flowchart of the proposed minimum processing NLE gain rule.}
  \label{fig:alg_flowchart}
\end{figure}
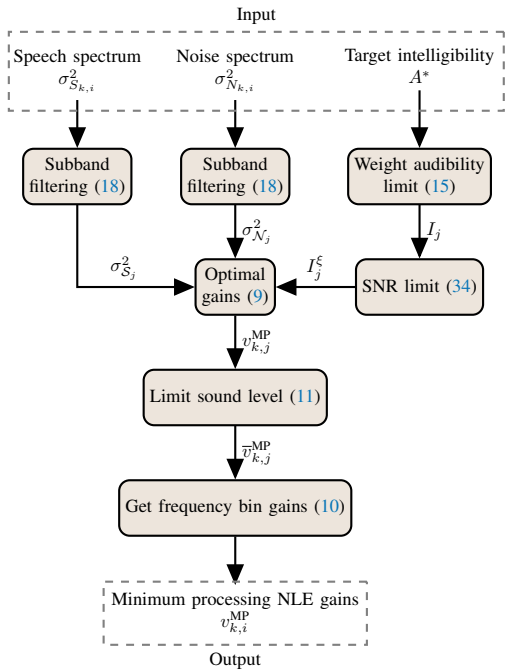

The proposed minimum processing NLE algorithm with ASII intelligibility criterion is summarized with the flowchart in Fig.~\ref{fig:alg_flowchart}.

  \section{Objective Performance Evaluation}\label{sec:objectiv_perform}
  In this section, we evaluate the performance of the proposed minimum processing NLE method~\eqref{eq:optimum_gains} in the NLE setup shown in Fig.~\ref{fig:nle_block}, where the target speech signal is processed prior to play out in the noisy environment, and the gain rule follows the proposed algorithm summarized in Fig.~\ref{fig:alg_flowchart}. We evaluate objective SI and SQ performance, and compare performance with existing related state-of-the-art NLE methods. For intelligibility scoring, we use the chosen performance criterion ASII~\cite{taal_optimal_2013} and the more general intelligibility predictor ESTOI~\cite{jensen_algorithm_2016}. The goal of minimum processing is to reproduce the clean unprocessed input speech signal as well as possible. Therefore, quality is estimated using PESQ~\cite{itu-t_recommendation_2007} and Segmental SNR (Seg-SNR)~\cite{hansen_effective_1998}.
  

\subsection{Experimental Setup}

The speech material used for the evaluation are English sentences each a few seconds in duration from the test set in the TIMIT-database~\cite{garofolo_timit_1993} sampled at $\SI{16}{kHz}$.
In each trial a speaker is selected randomly without replacement across trials. Afterwards, a random sentence is selected amongst those available for the selected speaker. Finally, the speech excerpt is padded with $\SI{0.5}{\s}$ of silence at the beginning and $\SI{0.125}{\s}$ of silence at the end.

For the noise signals we consider white noise, synthetic speech shaped noise, as well as noise recorded inside a car traveling at $\SI{130}{\km\per\hour}$. For the recorded noise, a random excerpt of the appropriate length is cut out of the noise recording for each trial.

Speech and noise signals were transformed to time-frequency domain using an STFT with $\SI{32}{ms}$ Hann windows with $50\%$ overlap at a sampling frequency of $\SI{16}{\kHz}$.
For the subbands, we consider a total of $J=30$ overlapping auditory filters with linearly spaced center frequencies on the equivalent rectangular bandwidth scale from \SIrange{150}{8000}{\Hz}\cite{van_de_par_perceptual_2005}, see Appendix~\ref{app:subband_def} for further details on the subband definition.

The target intelligibility is set to $A^*=0.7$, unless otherwise is stated. We choose this target, because an ASII value of $0.7$ corresponds to almost full intelligibility~\cite{van_kuyk_evaluation_2018}, and since any higher value requires significantly more processing (see also Section~\ref{sec:minproc_effect}).
The band intelligibility limits, $I_j$ are weighted according to \eqref{eq:weighted_I_lim} for all trials.

For the maximum subband gain, $v_{j,\text{max}}$, in \eqref{eq:max_gain}, the maximum allowed power $P^{\text{max}}_S$ is chosen such that
\begin{equation}
  10 \log \left(\frac{P^{\text{max}}_S}{P_0}\right) = 100\,\text{dB}_{\text{SPL}}.
\end{equation}
Here $P_0$ represents the digital reference power corresponding to a reference sound pressure~\cite{sauert_near-end_2014}. We make the convention that a signal power of $P_0$ corresponds to a signal with an $RMS=1$, thus
\begin{equation}
  P_0 = 10^{\frac{-100\,\text{dB}_{\text{SPL}}}{10}}.
\end{equation}

Finally, we conduct a total of $10$ trials and evaluate the performances in each trial. The average performance across the trials is then taken as the final score for each combination of noise, SNR and gain rule. 

\subsection{Reference methods}
We consider three reference methods that are similar to our work: $(1)$ The original ASII optimization of \cite{taal_optimal_2013} because we also optimize for ASII, $(2)$ the well known SII optimization of \cite{sauert_recursive_2010} that clips SNRs above and below a certain level and $(3)$ the very recent NoiseProp algorithm of \cite{niermann_listening_2021} because it has a simple gain-rule very similar to ours. The method of \cite{niermann_listening_2021} has an additional tuning parameter, $\rho_{NV} \in [0,1]$. It is mentioned in \cite{niermann_listening_2021}, that informal listening has shown $\rho_{NV} \in [0.5, 0.8]$ is a good choice. Therefore, we select the value $\rho_{NV}=0.7$ for all experiments.

All reference methods are implemented following the procedure described in the previous subsection. The only difference is the applied gain rule. That is, for ease of comparison all methods are implemented as time-invariant even though the original work may have considered slowly time-varying processing, e.g.,~\cite{sauert_recursive_2010, niermann_listening_2021}.

All the reference methods are based on a target speech power equality constraint. Hence, they are designed to keep the total power across frequency at a certain level, $P_{\text{ref}}$. However, the proposed method is able to increase the power as much as needed, achieving a total processed power level denoted by $P_{\text{MP}}$. Therefore, for a fair comparison, the reference methods are also allowed to increase the power to that level. That is, the reference methods are implemented such that $P_{\text{ref}}=P_{\text{MP}}$.

\subsection{Minimum Processing Effect}\label{sec:minproc_effect}

Initially, we illustrate the effects of the proposed minimum processing solution in terms of processing penalty versus intelligibility performance and resulting increases in speech power. The effects are considered using a white noise source to limit the importance of the spectral distribution of noise.

\begin{figure}[!t]
  \centering
  \includegraphics[width=\columnwidth]{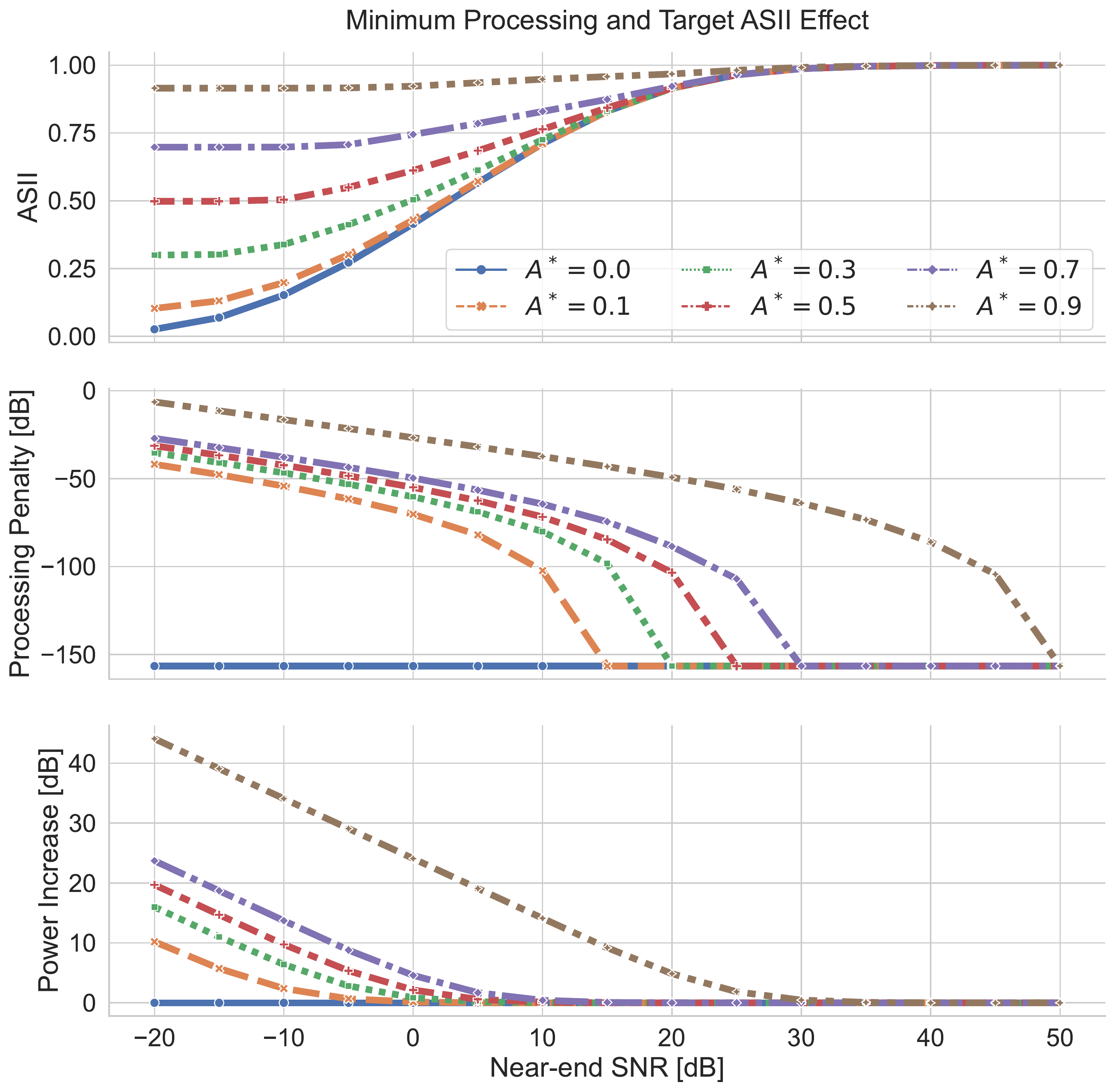}
  \caption{The achieved ASII (top), and the corresponding MSE processing penalty (middle) and power increase (bottom) versus input near-end SNR for varying choices of target intelligibility. The legend in the top panel applies to all plots.}
  \label{fig:minproc_effect}
\end{figure}

Fig.~\ref{fig:minproc_effect} shows that for increasing values of the target intelligibility, $A^*$, the achieved ASII (top) along with the corresponding MSE processing penalty (middle) and increase in speech power (bottom). The lower limit case of target AII $A^*=0$ corresponds to the case of no processing. The ASII is upper bounded by $A^*\leq1$ and is only achievable in the limit case of infinite SNR, therefore we only show the trend up to $A^*=0.9$.
The top panel in Fig.~\ref{fig:minproc_effect} shows, that when the target intelligibility $A^* > 0$, we are able to improve the performance over the unprocessed case ($A^*=0$) for a large range of SNRs. From the middle and bottoms panels in Fig.~\ref{fig:minproc_effect}, we see that the greater the target intelligibility and the lower the input SNR, the increases in intelligibility performance come at the cost of a greater need for processing and increased speech power. Finally, as the near-end SNR improves, the target intelligibility is more easily achieved and the processing required to achieve a given target decreases, as can be seen by a transition from a high level of processing to no processing.

\subsection{Objective Performance}\label{sec:obj_score}

Fig.~\ref{fig:obj_score_car_SSN} shows the ASII, ESTOI, PESQ and Seg-SNR performance as function of the near-end SNR for the proposed minimum processing NLE and the reference methods in both car noise and speech shaped noise. Further experiments using cafeteria and babble noise sources show performance similar to experiments using speech shaped noise, however these are omitted here do to space limitations.

\subsubsection{Estimated intelligibility performance}

From the SI metrics ASII and ESTOI, we see that the proposed method improves SI over the unprocessed speech to the desired level at the lower SNRs. The proposed method even achieves the best ESTOI performance in car noise compared to the reference methods. In ASII, the reference methods are able to gain a slightly better SI than the proposed. This is expected, as they try to maximize the SI given the allowed power constraint and not just reach a sufficient level of SI.
As the SNR increases and the speech becomes naturally intelligible, the proposed method follows the unprocessed performance above the target intelligibility. Thus, the proposed method always achieves the desired SI performance or better. Similar performance can be seen with SII maximization by Sauert~\cite{sauert_recursive_2010}, however this does not happen until the SII clips the subband SNRs above $\SI{15}{\dB}$ and the SII is fully maximized. Additionally, as the unprocessed SI reaches its ceiling the reference methods are no longer able to further enhance SI by maximization.

\subsubsection{Estimated speech quality performance}
Generally, from the PESQ and Seg-SNR plots, we see that all Seg-SNR and PESQ scores are at the lowest levels for very low SNRs, and it is difficult to tell any difference in quality performances. This is expected, since the near-end SNR is low and the noise is the dominant part of the signals. 
As the input SNR increases so does the PESQ and Seg-SNR scores, where the unprocessed clean speech is considered to be the maximum level of SQ. This is due to PESQ and Seg-SNR measuring quality by how similar the signal presented to the listener, $Z$, is to the clean reference, $S$. Hence, for high SNR this is the maximum achievable level when the near-end noise is insignificant and $Z \approx S$.

For increasing SNRs in both car noise and speech shaped noise, we see that the proposed method follows the maximum SQ performance of the unprocessed signal, when the intelligibility is at the desired level and the processing switches off. Similarly, the method of Sauert~\cite{sauert_recursive_2010} switches off automatically at maximum SII, but does so at a higher level of intelligibility, thereby incurring more speech distortions than the proposed method. On the other hand, the reference methods of Taal~\cite{taal_optimal_2013} and Niermann~\cite{niermann_listening_2021} never switch off and continue processing at the higher SNRs causing large speech distortions.

Comparing between car noise and speech shaped noise we see that the PESQ and Seg-SNR show less speech distortion in speech shaped noise. Further, since car noise is a less severe noise, speech distortions occur at lower SNRs and are more clearly heard in car noise. Thus, excessive processing is more detrimental to SQ in better noise conditions.

\begin{figure*}[!t]
  \centering
  \subfloat[]{\includegraphics[width=\columnwidth]{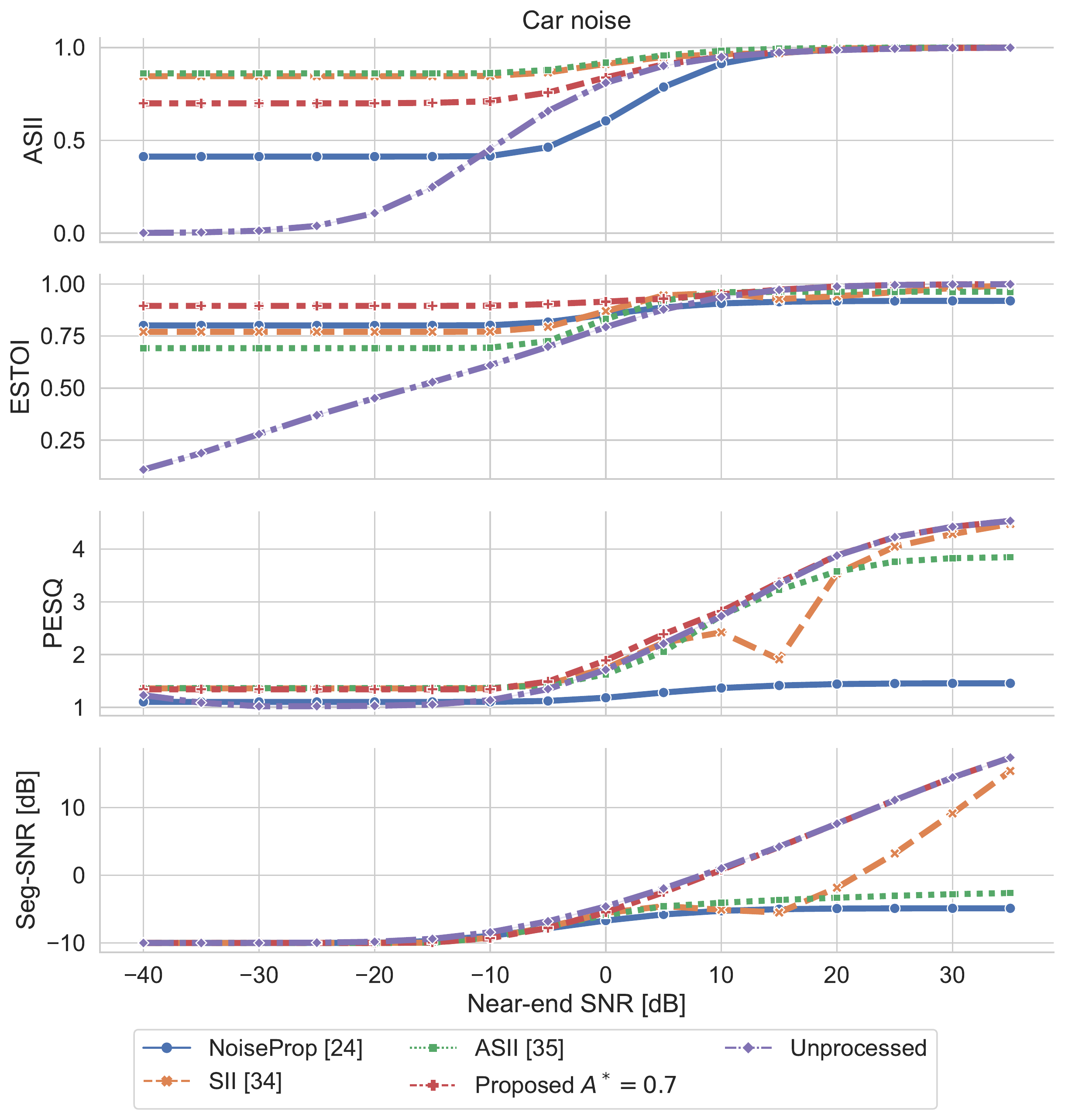}%
  \label{fig:obj_score_car}}
  \hfil
  \subfloat[]{\includegraphics[width=\columnwidth]{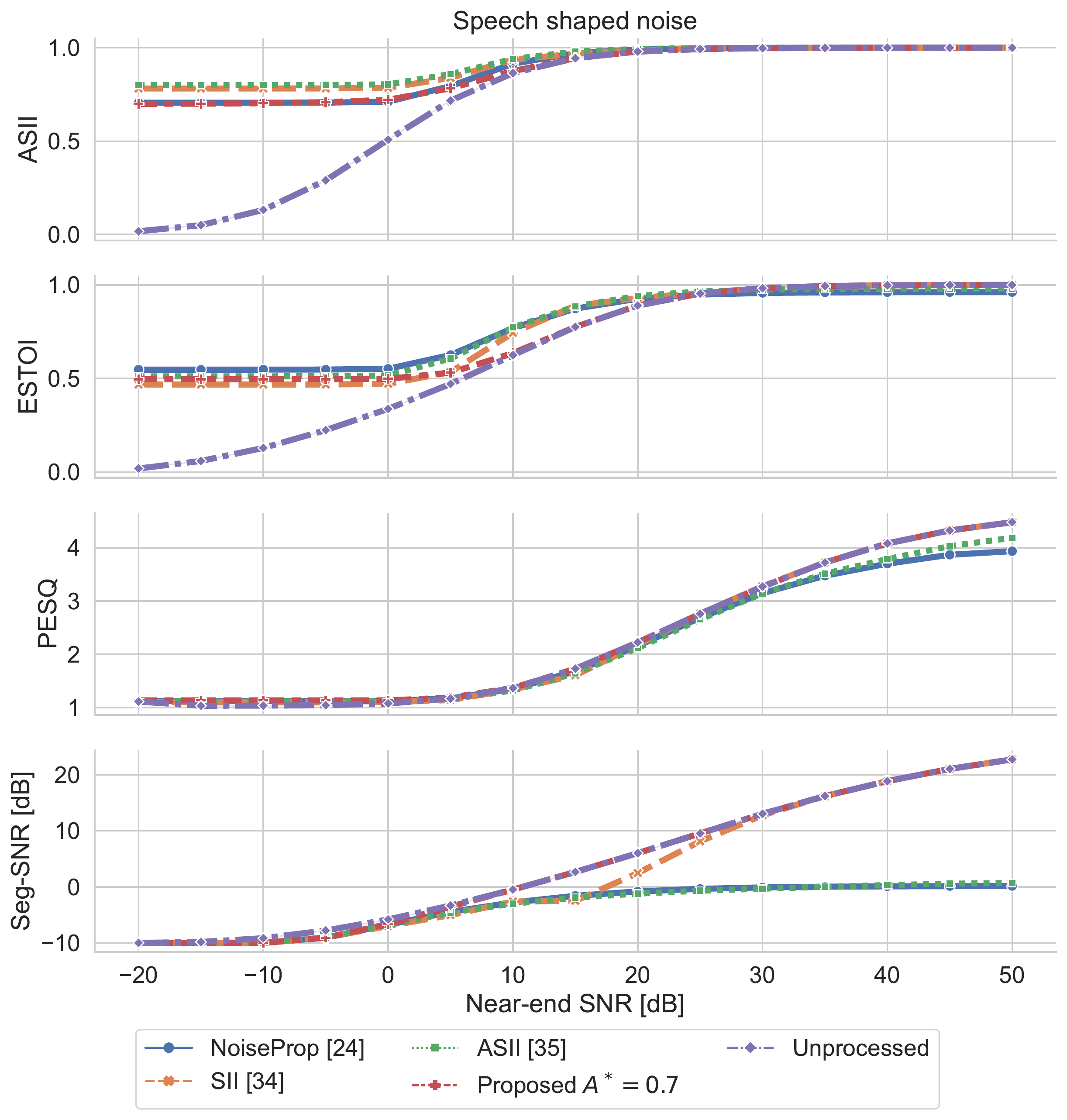}%
  \label{fig:obj_score_SSN}}
  \caption{ASII, ESTOI, PESQ and Seg-SNR scores in (a) car noise and (b) speech shaped noise. Legends apply to all plots within a column.}
  \label{fig:obj_score_car_SSN}
  \end{figure*}

\subsection{Gain dynamics}
To show the advantages of minimum processing in comparison with existing methods, we illustrate how the proposed method and \cite{niermann_listening_2021} affect the speech spectra in different noise conditions. We consider the behavior in car noise for an input SNR of $\SI{-30}{\dB}$ and $\SI{10}{\dB}$. In these SNRs both the proposed method and \cite{niermann_listening_2021} achieve very high SI, however they have a vastly different SQ at $\SI{10}{\dB}$ as seen in Fig.~\ref{fig:obj_score_car}.

\begin{figure}[!t]
  \centering
  \includegraphics[width=\columnwidth]{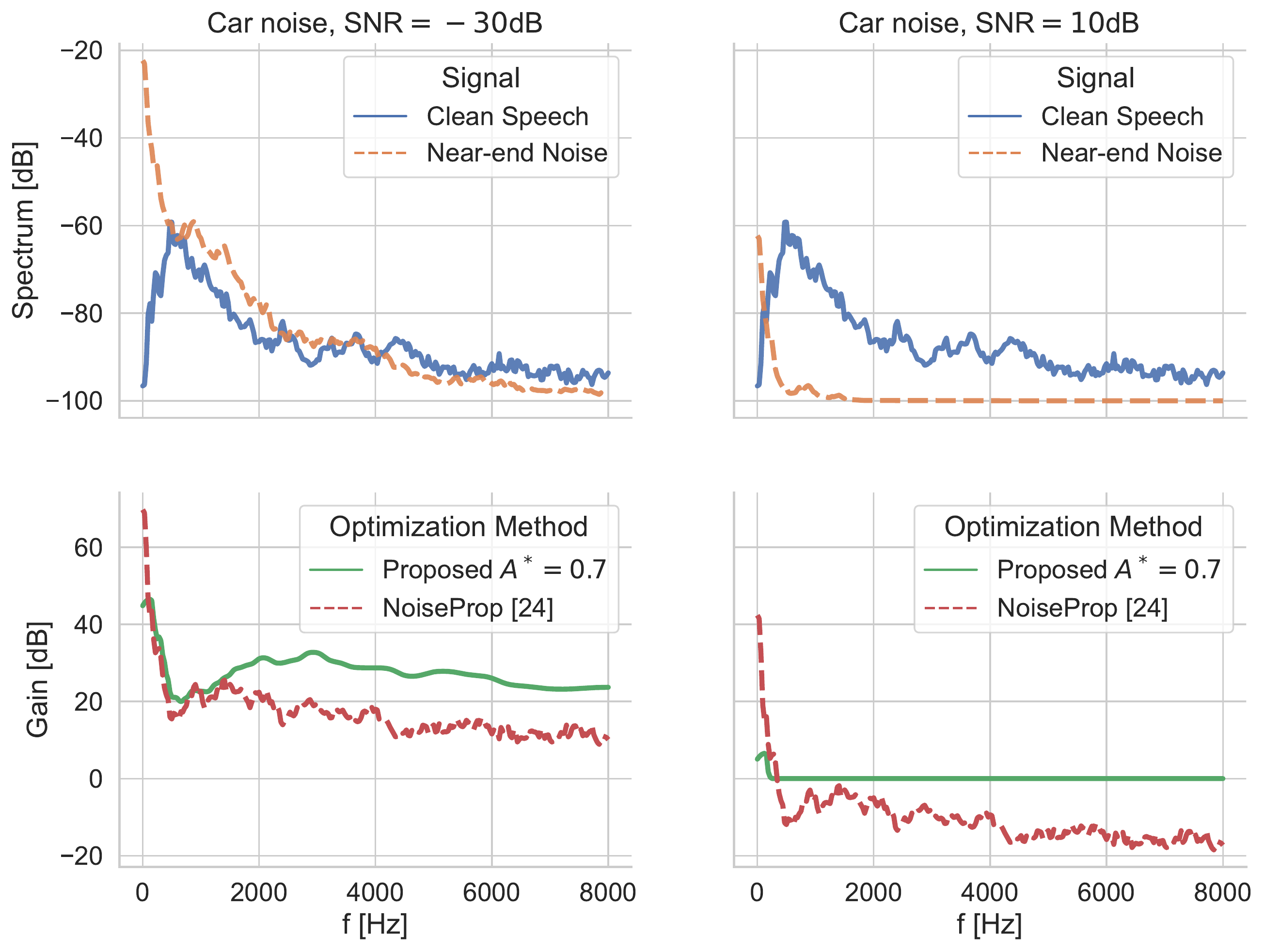}
  \caption{Clean speech and near-end noise spectra (top row), and optimal gains (bottom row) for car noise in $\SI{-30}{\dB}$ SNR (left column) and $\SI{10}{\dB}$ SNR (right column).}
  \label{fig:spectrum_effect}
\end{figure}

Fig.~\ref{fig:spectrum_effect} shows an example of the clean speech and noise spectrum (top row), as well as the derived optimal gains for both methods (bottom row) for both $\SI{-30}{\dB}$ SNR (left column) and $\SI{10}{\dB}$ SNR (right column). Looking at the left-hand column with low SNR, we see that the proposed method provide optimal gains that raise the speech power sufficiently above the near-end noise, and the reference method~\cite{niermann_listening_2021} optimal gains are proportional to the near-end noise. Now considering the difference between the two SNRs (columns), we see how the proposed method automatically minimizes the amount of processing as the SNR increases and the SI is at the ceiling level. This results in high SI at low SNRs, and both high SI and SQ at high SNRs in Fig.~\ref{fig:obj_score_car}. On the other hand, the reference~\cite{niermann_listening_2021} method, which always tries to maximize SI, continues to process the speech signal proportionally to the near-end noise even at the high SNR when SI is at the ceiling level. This continued processing introduces audible speech distortions, which result in low SQ scores at the higher SNRs as seen in Fig.~\ref{fig:obj_score_car}.

\section{Subjective Speech Quality Test}\label{sec:subjec_qual_test}
In this section, we evaluate the performance of the proposed minimum processing NLE method~\eqref{eq:optimum_gains} by a subjective listening test for SQ.
Speech quality and general audio quality is highly subjective and may not always be represented well by objective measures. For example, grading SQ in the presence of noise using PESQ is difficult due to a high sensitivity to the environmental noise compared to the speech distortion~\cite{zahedi_minimum_2021}. 
Therefore, to 
further investigate the performance of the proposed procedure we perform a listening test to asses the SQ in the near-end listening scenario, where we cannot process the noise but only the clean speech component. 
We consider the proposed minimum processing in comparison with the highly similar and state-of-the-art NoiseProp method of \cite{niermann_listening_2021}. We investigate performance in both car noise and speech shaped noise.

\subsection{Listening Test Setup}
The listening test was conducted by $21$ ($4$ female, $17$ male) volunteering untrained listeners. The participants had an age span of $22$ to $59$ years with an average age of $39.6$ years. All participants had self-reported normal hearing. The average test time including training was $38$\,minutes.

The listening test was conducted in a silent room using a Lenovo~T$570$ laptop equipped with a USB sound card (DragonFly Black) and a pair of closed headphones (Beyerdynamic DT-$770$ Pro $32$\,ohm) for reporting and audio playback. 
The user interface was based on the Web Audio Evaluation Tool~\cite{jillings_web_2015}. All audio stimuli were normalized to the same perceived loudness of $-30$\,LUFS following  \cite{itu-r_recommendation_2015-1} and the EBU R$128$~\cite{ebu_ebu_2014} recommendation for loudness normalization as implemented within the Web Audio Evaluation Tool~\cite{man_web_2022}. Participants were allowed to adjust the general volume to a comfortable level during the training session of the test.

\subsection{Procedure}
We conducted a listening test following the MUlti Stimulus with Hidden Reference and Anchor (MUSHRA)~\cite{itu-r_recommendation_2015} paradigm, where the audio quality is assessed on a scale from $0$ to $100$, divided into five equal intervals labelled as \textit{bad, poor, fair, good} and \textit{excellent}. The participants were instructed to grade the \textit{basic audio quality} compared to a known reference signal. No other definition was provided of audio quality.

Each test participant was presented with $2$ sequences of $8$ trials, one sequence for each of the car and speech shaped noise. Each trial consisted of a reference signal (unprocessed signal in high SNR) and five other signals to be rated: $1$~hidden reference, $2$~systems under test (proposed method and NoiseProp~\cite{niermann_listening_2021}), $1$~unprocessed signal, and $1$~hidden anchor (unprocessed signal at lower SNR).
The SNRs considered for the reference, anchor and test cases vary depending on the noise type, such that the perceptual quality was not affected too severely by a loss in intelligibility. For car noise, the reference and anchor signal SNRs were set to $\SI{15}{\dB}$ and $\SI{-30}{\dB}$ respectively. For speech shaped noise, the reference and anchor signal SNRs were set to $\SI{25}{\dB}$ and $\SI{-5}{\dB}$ respectively. For both noise types, four trials were used to evaluate the systems at an input SNR of $\SI{0}{\dB}$. The last $4$ trials were used to evaluate the systems at $\SI{-10}{\dB}$ and $\SI{10}{\dB}$ for car noise and speech shaped noise, respectively. 

Before the actual experiment, each participant went through a short training session, where they were able to get used to the task at hand, the different audio stimuli and the user interface. The sentences in the training were different from those in the actual experiment and did not contribute to the final test score.

\subsection{Listening Test Results and Discussion}
\begin{figure}[!t]
  \centering
  \includegraphics[width=\columnwidth]{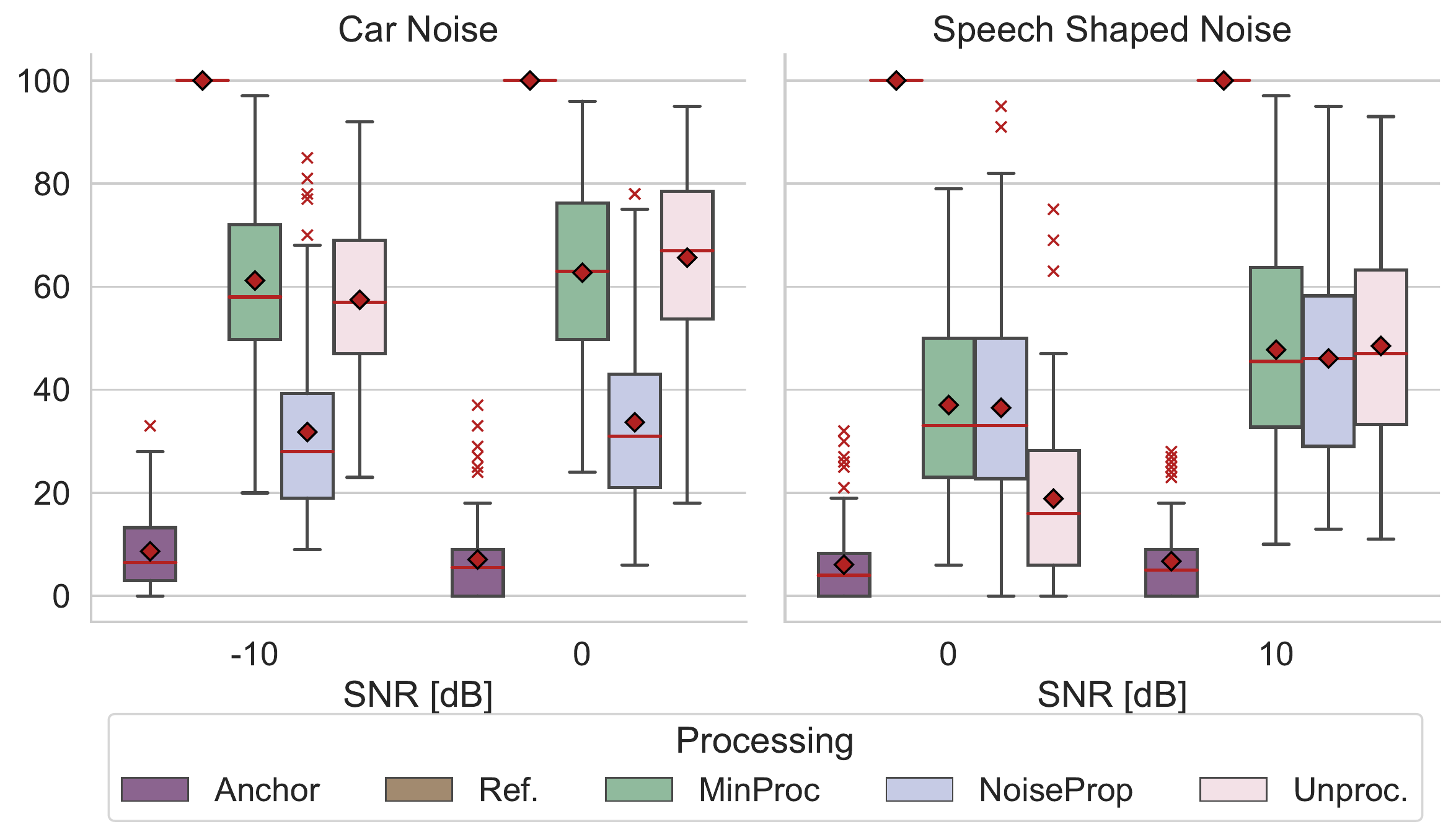}
  \caption{Boxplot illustration of the MUSHRA listening test results for car noise (left) and speech shaped noise (right). Medians and means are indicated by red horizontal lines and diamonds, respectively. Outliers (according to the $1.5$ interquantile range rule) are indicated by red crosses.}
  \label{fig:listening_test_boxplot}
\end{figure}

The average scores across trials and participants for each noise, processing and SNR condition are shown in Fig.~\ref{fig:listening_test_boxplot} with boxplots. 

The speech shaped noise results at an SNR of $\SI{0}{\dB}$ in Fig.~\ref{fig:listening_test_boxplot}, show that processing with the proposed method and NoiseProp \cite{niermann_listening_2021} to increase intelligibility has a positive effect on subjective quality compared to the unprocessed performance, when the noise situation is more severe. As the SNR increases to $\SI{10}{\dB}$, we cannot detect any effect on the subjective SQ compared to the unprocessed performance. This is in contrast to the PESQ results in Fig.~\ref{fig:obj_score_SSN}, where no difference was seen.

The listening test results for car noise at both $\SI{-10}{\dB}$ and $\SI{0}{\dB}$ SNR show that NoiseProp~\cite{niermann_listening_2021} deteriorates SQ, while the proposed minimum processing does not. Both methods have comparable SI performance at these SNRs with the proposed method slightly outperforming the NoiseProp, as shown by ESTOI in Fig.~\ref{fig:obj_score_car}. The SQ results are also confirmed with PESQ at $\SI{0}{\dB}$ SNR in Fig.~\ref{fig:obj_score_car}, where there is a score difference of almost $2$ points. At input SNR of $\SI{-10}{\dB}$ PESQ only shows a minor difference between the two methods.
This illustrates, when the unprocessed SI is almost maximized due to favorable noise conditions, the excessive processing of NoiseProp, as seen Fig.~\ref{fig:spectrum_effect}, can harm subjective SQ, even though the SI may be increased slightly. 

These observations confirm the results of the objective SQ metrics as well as those in the literature~\cite{tang_study_2018}, that SI enhancement may lead to both decreases and increases in SQ. Particularly, decreases in SQ are observed when the unprocessed intelligibility is already high and the processing becomes excessive. On the other hand, in harsh noise conditions, the environmental noise masks most speech distortions. This is also in line with the results of \cite{tang_study_2018}, where SI is shown to be the dominant factor for SQ in severe noise.


Existing NLE methods focus on maximizing SI. Therefore, given a certain allowed increase in speech power, the reference methods are sometimes able to gain a slightly higher SI than the proposed method as the ASII and ESTOI scores show in Fig~\ref{fig:obj_score_car_SSN}. 
However, 
the natural ceiling effect of SI occurs prior to the maximum value of the metric~\cite{van_kuyk_evaluation_2018}. Therefore, the objective SQ and listening test results also show the achieved SI score may be unnecessarily high and lead to excessive speech distortion. 
Instead, by using a single intelligibility target, $A^*$, our proposed method and can be better linked to the onset of the ceiling effect for various noise types. Thus, by varying the target intelligibility, $A^*$, we see how the proposed method allows controlling the tradeoff between SI and SQ.

\section{Conclusion}\label{sec:conclusion}

We proposed a (novel) near-end listening enhancement concept, where the output signal is optimized to have the minimum amount of processing artifacts (for example, distance from the clean speech signal), with the constraint that a certain performance criterion (for example, estimated intelligibility) is satisfied. The proposed concept is adaptive to environmental noise conditions by focusing on estimated intelligibility for high noise levels, and, as a consequences of the minimum processing paradigm, quality for low noise levels. The trade-off between intelligibility and quality can be controlled via the performance criterion.
To demonstrate an instance of the proposed framework, we make a thorough investigation of an example case, where the performance criterion is based on the approximated speech intelligibility index and the processing metric is the mean squared error. We show that, this provides a simple gain rule based NLE, that raises the speech spectrum above the noise to exactly the minium required level for the desired intelligibility. The proposed method, is able to distribute the processing across subbands to maximize or reach a target intelligibility by varying the desired intelligibility target between the subbands.
If the noise conditions are favorable, the processing is automatically reduced leading to fewer processing artifacts and hence improved speech quality. Thereby, by allowing an increase in speech power, the proposed NLE technique is able to adaptively achieve any desired intelligibility level  with a minimum of processing artifacts. 
Experimental studies verify the advantages of the concept in terms of both intelligibility and quality.

Future work will include expanding the proposed concept to jointly consider near-end listening enhancement together with far-end noise reduction for an increased intelligibility and quality performance. Additionally, it is interesting combining the proposed method with acoustic echo cancellation techniques when working with joint far-end and near-end optimization.

\appendices

\section{Subband Weights}\label{app:subband_def}

Assigning frequency bins to subbands and determining the weight of their contributions can be done in different ways, cf.~\cite[App. A]{zahedi_minimum_2021}. We consider a gammatone filter bank model~\cite{van_de_par_perceptual_2005} that we normalize to preserve power between subband and frequency domain.


We focus on subbands in terms of overlapping auditory filters. Let $h_j$ be the impulse response of the $j$'th subband auditory filter. Then, the energy of the clean speech signal, $\mathcal{S}_{j,i}$, is given as the convolution with between $s$ and $h_j$, which in in time-subband domain is given as
\begin{equation}
  \mathcal{S}_{j,i}^2 \triangleq \sum_{k \in \mathbb{B}_j} \lvert S_{k,i} \rvert^2 
  \lvert H_j(k) \rvert^2 ,
\end{equation}
where $H_j(k)$ represents the DFT of $h_j$ in frequency-bin $k$. We consider squared magnitude responses, $\lvert H_j(k) \rvert^2$ given by the gammatone filter bank derived in \cite{van_de_par_perceptual_2005}. Furthermore, we want to ensure the total power is the same in both the frequency and subband domain, i.e.,
\begin{equation}
    \sum_j \sigma_{\mathcal{S}_{j,i}}^2 = \sum_k \sigma_{S_{k,i}}^2. \label{eq:cb_dft_ener_eq}
\end{equation}
Now inserting the filtering operation we have that
\begin{align}
    \sum_j  \sigma_{\mathcal{S}_{j,i}}^2 &= \sum_j \sum_k \lvert H_j(k)\rvert^2\sigma_{S_{k,i}}^2\\
&=\sum_k \sum_j \lvert H_j(k)\rvert^2\sigma_{S_{k,i}}^2. \label{eq:cb_filt_doub_sum}
\end{align}
From \eqref{eq:cb_filt_doub_sum} we see that \eqref{eq:cb_dft_ener_eq} is satisfied if $\sum_j \lvert H_j(k)\rvert^2 = 1$. We can achieve this by normalizing $H_j(k)$, i.e.,
\begin{equation}
    H'_j(k) = \frac{H_j(k)}{\sqrt{\sum_l^J \lvert H_l(k)\rvert^2}}, \quad \forall j.
\end{equation}
%
Hence, we let the subband filter weights, $\omega_{j,k}$, be the normalized squared magnitude response of $h_j$, i.e., $\omega_{j,k}= \lvert H'_j(k) \rvert^2$.

\section{MSE Processing Penalty}\label{app:mse_proc_penalty}
For the $j$'th subband we write the MSE processing penalty as
\begin{align}
    \mathcal{D}_j(\bm{S}_j, \bm{Z}_j) &=
     \sum_{k\in \mathbb{B}_j}\omega_{j,k} \operatorname{E}\left[\lvert S_{k} - Z_{k}\rvert^2\right]\\
     &=\sum_{k\in \mathbb{B}_j}\omega_{j,k}\left(\left( 1 - v_{k} \right)^2 \sigma_{S_{k}}^2 + \sigma_{N_{k}}^2 \right),
\end{align}
where we have assumed the STFT coefficients are uncorrelated across frequency.
Thereby, the MSE in a subband is the `weighted average' of the MSE in the DFT domain where the weights are the subband filter weights. Since the near-end noise power, $\sigma_{N_{k}}^2$, is unaffected by the processing, $v_{k}$, we can disregard it from the processing penalty, and we have
\begin{equation}
  \mathcal{D}_j(\bm{S}_j, \bm{Z}_j)=\sum_{k\in \mathbb{B}_j}\omega_{j,k}\left( 1 - v_{k} \right)^2 \sigma_{S_{k}}^2.
\end{equation}

\section{ASII performance criterion}\label{app:asii_perform_criterion}
The ASII~\cite{taal_optimal_2013} is defined as
 \begin{equation}
     ASII \triangleq \sum_{j}\gamma_j f(\xi_j),
 \end{equation}
where the weights $\gamma_j$ are pre-defined constants (more specifically the band importance functions as defined in \cite{american_national_standards_institute_methods_2017}),
\begin{equation}
  f(\xi_j)\triangleq \frac{\xi_j}{\xi_j + 1},\label{eq:sii_apx_fxi}
\end{equation}
is the sigmoidal audibility function per subband,
and $\xi_j$ is the SNR per subband, i.e.,
\begin{equation}
  \xi_j \triangleq \frac{\sigma_{\widetilde{\mathcal{S}}_j}^2}{ 
     \sigma_{\mathcal{N}_j}^2
}, \label{eq:full_snr} 
\end{equation}
where 
\begin{equation}
    \sigma_{\widetilde{\mathcal{S}}_j}^2 =  \sum_{k\in \mathbb{B}_j}\omega_{j,k} 
    v_{k}^2 \sigma_{S_{k}}^2,\label{eq:proc_subband_power}
\end{equation}
is the processed speech power. Hence, the \textit{processed} subband SNR is the ratio of the \textit{processed} speech power to the near-end noise power at the listener.

%
%
%
%

Let $I_j$ be a given minimum requirement on the audibility performance in a particular subband, $j$, i.e., we require 
 \begin{equation}
   f(\xi_j) \geq I_j. \label{eq:audibility_req}
 \end{equation}
It is beneficial to interpret the subband audibility limits, $I_j$, in terms of the subband SNR.
Using \eqref{eq:sii_apx_fxi}, it follows that \eqref{eq:audibility_req} can be written as
\begin{align}
     \xi_j
     &\geq \frac{I_j}{1-I_j} \triangleq I_j^\xi\label{eq:SNR_lim_def}
  \end{align}
where $I_j^\xi \triangleq \frac{I_j}{1-I_j}$, may be considered a desired minimum processed SNR. 
Therefore, instead of focusing on choosing an intelligibility limit, $I_j$, we can equivalently choose an SNR limit $I_j^\xi$, or interpret the desired audibility limits as a limit on the subband SNR. Hence, the performance criterion in \eqref{eq:gen_minproc_problem} is chosen as the subband SNR, i.e.,
\begin{equation}
  \mathcal{I}_j\left(\bm{S}_j, \bm{Z}_j\right)= \xi_j= \frac{\sigma_{\widetilde{\mathcal{S}}_j}^2}{ 
       \sigma_{\mathcal{N}_j}^2
  }.
\end{equation}
Writing out the performance criterion we have that 
\begin{align}
  \mathcal{I}_j\left(\bm{S}_j, \bm{Z}_j\right)&\geq I_j^\xi\\
  \sum_{k\in \mathbb{B}_j} \omega_{j,k}
         v_{k}^2 \sigma_{S_{k}}^2  &\geq \sigma_{\mathcal{N}_{j}}^2 I_j^\xi.
\end{align}

\section{Proof of Theorem~\ref{theo:minproc_solution}}\label{app:opt_sol_deriv}
We first see that the optimization problem \eqref{eq:gen_minproc_problem} can be written as \eqref{eq:NLE_minproc_problem} by inserting \eqref{eq:mse_penal} and \eqref{eq:SNR_crit} in \eqref{eq:gen_minproc_problem}. 

To prove that the solution to \eqref{eq:NLE_minproc_problem} is given by the weights in \eqref{eq:optimum_gains}, we start by making an observation on the relationship between the MSE cost function and the performance constraint in \eqref{eq:NLE_minproc_problem}. Clearly, the unconstrained solution to the MSE minimization problem is $v_k = 1$ for $k \in \mathbb{B}_j$, i.e., no processing obviously minimizes the MSE. 

The interesting scenario is then that of ${\sum_{k\in \mathbb{B}_j}\omega_{j,k}  \sigma_{S_{k}}^2  < \sigma_{\mathcal{N}_{j}}^2 I_j^\xi}$. Since we have no energy constraint and we must increase the speech energy to not violate the performance constraint we can lower bound the gains as $v_k \geq 1$ for all $k\in \mathbb{B}_j$, i.e., we only increase the speech power and have no reason to decrease it. Hence, the optimization problem becomes 
\begin{equation}
    \begin{array}{ll}
        \displaystyle \min_{\{v_k\} \in \mathbb{R}_+, k\in\mathbb{B}_j} & \sum_{k\in \mathbb{B}_j}\omega_{j,k}\left( 1 - v_{k}\right)^2 \sigma_{S_{k}}^2,\\
        \mbox{subject to} &  \sum_{k\in \mathbb{B}_j}\omega_{j,k} 
         v_{k}^2 \sigma_{S_{k}}^2  \geq \sigma_{\mathcal{N}_{j}}^2 I_j^\xi,\\
         &v_k \geq 1.
    \end{array}
\end{equation}
Formulating the Lagrangian we have 
\begin{align*}
    \mathcal{L}(\{v_k\}, \lambda_j, \{\mu_k\}) &= \sum_{k\in \mathbb{B}_j}\omega_{j,k}\left( 1 - v_{k}\right)^2 \sigma_{S_{k}}^2 + \mu_k (1-v_k) \nonumber\\
   &\quad + \lambda_j \left(
        \sigma_{\mathcal{N}_{j}}^2 I_j^\xi - \sum_{k\in \mathbb{B}_j}\omega_{j,k} v_{k}^2 \sigma_{S_{k}}^2 \right)
\end{align*}
Taking the derivative with respect to $v_k$ for a particular ${k \in \mathbb{B}_j}$, we have
\begin{align}
    \frac{\partial}{\partial v_k}\mathcal{L} &= -2\omega_{j,k}\left( 1 - v_{k}\right) \sigma_{S_{k}}^2
    - 2\lambda_j\omega_{j,k} v_{k} \sigma_{S_{k}}^2 - \mu_k.
\end{align}
Since we optimize over a non-convex set the optimization problem is non-convex. Thus, the Karush-Kuhn-Tucker conditions for this problem are not sufficient for a global optimum. However, the they are still necessary conditions, thus we use these to determine a solution to the problem.
\begin{subequations}
  \begin{align}
    v_k \geq 1, \quad
    \mu_k \geq 0, \quad
    \mu_k(1- v_k) = 0, \quad
    \lambda_j &\geq 0\\
    \sigma_{\mathcal{N}_{j}}^2 I_j^\xi -    \sum_{k\in \mathbb{B}_j}\omega_{j,k} 
    v_k^2 \sigma_{S_{k}}^2 &\leq 0\\
    \lambda_j \left[ \sigma_{\mathcal{N}_{j}}^2 I_j^\xi - \sum_{k\in \mathbb{B}_j}\omega_{j,k} 
    v_k^2 \sigma_{S_{k}}^2 \right] &= 0\label{eq:KKT_ne_only_perform_con}\\
    -2\omega_{j,k}\left( 1 - v_{k}\right) \sigma_{S_{k}}^2
    - 2\lambda_j\omega_{j,k} v_{k} \sigma_{S_{k}}^2 - \mu_k &= 0\label{eq:KKT_ne_only_stat}
  \end{align}
\end{subequations}
Isolating $\mu_k$ in the last equation we have,
\begin{align}
    -2\omega_{j,k}\left( 1 - v_{k}\right) \sigma_{S_{k}}^2- 2\lambda_j\omega_{j,k} v_{k} \sigma_{S_{k}}^2 \geq 0.
\end{align}
Now solving for $v_k$,
\begin{align}
    v_k &\geq \frac{1}{1 - \lambda_j}
\end{align}
Combining this with $v_k \geq 1 $, we have
\begin{equation}
    v_k = \max \left(1, \frac{1}{1 - \lambda_j}\right).\label{eq:ne_only_gain_max}
\end{equation}
We observe that $v_k=v_{k'}$ for all $k,k' \in \mathbb{B}_j$. That is, the gains are equal for all frequencies within subband $j$.

Now to determine $\lambda_j$, we first notice that if $\lambda_j = 0$ then $v_k = 1~\forall k \in \mathbb{B}_j$. Then the performance constraint can only be satisfied if $ \sigma_{\mathcal{S}_{j}}^2 \geq \sigma_{\mathcal{N}_{j}}^2 I_j^\xi$, i.e., if we are in the scenario that does not require processing.
Therefore, we must have ${\lambda_j > 0}$ if $ \sigma_{\mathcal{S}_{j}}^2  < \sigma_{\mathcal{N}_{j}}^2 I_j^\xi$. Furthermore, from \eqref{eq:ne_only_gain_max} we see that the non-trivial solution requires $\lambda_j < 1$. Therefore, \eqref{eq:KKT_ne_only_perform_con} is only satisfied for $\lambda_j \in (0,1)$ if the optimal $\lambda_j$ satisfy
\begin{align}
    \sigma_{\mathcal{N}_{j}}^2 I_j^\xi -    \sum_{k\in \mathbb{B}_j}\omega_{j,k}  \frac{1}{\left(1 - \lambda_j\right)^2}\sigma_{S_{k}}^2 &= 0\\
     \lambda_j &= 1 - \sqrt{\frac{\sigma_{\mathcal{S}_{j}}^2 }{\sigma_{\mathcal{N}_{j}}^2 I_j^\xi}}.
\end{align}
%
Inserting this into \eqref{eq:ne_only_gain_max} the optimal gain is
\begin{align}
    v_k^* 
     &= \max \left(1, \sqrt{\frac{\sigma_{\mathcal{N}_{j}}^2 I_j^\xi}{\sigma_{\mathcal{S}_{j}}^2}}\right)
     = \sqrt{\frac{\sigma_{\mathcal{N}_{j}}^2 I_j^\xi}{\sigma_{\mathcal{S}_{j}}^2}},
\end{align}
where the second equality follows since we consider the case of $\sigma_{\mathcal{N}_{j}}^2 I_j^\xi > \sigma_{\mathcal{S}_{j}}^2$. Finally, by inserting the optmium $v_k^*, \lambda_j^*$ and $\mu_k^*$ in \eqref{eq:KKT_ne_only_stat}, it can be seen that this is a stationary point of the Lagrangian and thus dual feasible. 
Hence, the solution is
\begin{equation}
  v_{k}^{*} = \begin{cases}
    1 & \text{if}~\sigma_{\mathcal{S}_{j}}^2  \geq \sigma_{\mathcal{N}_{j}}^2 I_j^\xi\\
    \sqrt{\frac{\sigma_{\mathcal{N}_{j}}^2 I_j^\xi}{\sigma_{\mathcal{S}_{j}}^2}}  & \text{otherwise}
  \end{cases}, \quad \forall k \in \mathbb{B}_j.
\end{equation}


This completes the proof.

\bibliographystyle{IEEEtran}
\bibliography{IEEEabrv,NLE_minproc_paper_bib_new.bib}

\vskip -2\baselineskip plus -1fil
\begin{IEEEbiographynophoto}{Andreas Jonas Fuglsig}
  (S'22) received the B.Sc. and M.Sc. degrees in mathematical engineering from Aalborg University in 2017 and 2019, respectively. He joined RTX A/S as an R\&D engineer in 2019. Since 2020, he has been working towards the Ph.D. degree with the Centre on Acoustic Signal Processing Research (CASPR) at Aalborg University and RTX A/S. His research interests include acoustic signal processing, speech enhancement and information theory.
\end{IEEEbiographynophoto}

\vskip -2\baselineskip plus -1fil
\begin{IEEEbiographynophoto}{Jesper Jensen}
  received the M.Sc. degree in electrical
  engineering and the Ph.D. degree in signal processing
  from Aalborg University, Aalborg, Denmark,
  in 1996 and 2000, respectively. From 1996 to 2000,
  he was with the Center for Person Kommunikation
  (CPK), Aalborg University, as a Ph.D. student and
  Assistant Research Professor. From 2000 to 2007, he
  was a Post-Doctoral Researcher and Assistant Professor
  with Delft University of Technology, Delft,
  The Netherlands, and an External Associate Professor
  with Aalborg University. Currently, he is a Fellow at Oticon A/S, Denmark, 
  where his main responsibility is scouting and development of new signal processing concepts
  for hearing aid applications. He is a Professor with the Section for Signal
  and Information Processing (SIP), Department of Electronic Systems, at
  Aalborg University. He is also a co-founder of the Centre for Acoustic Signal
  Processing Research (CASPR) at Aalborg University. His main interests are
  in the area of acoustic signal processing, including signal retrieval from
  noisy observations, coding, speech and audio modification and synthesis,
  intelligibility enhancement of speech signals, signal processing for hearing
  aid applications, and perceptual aspects of signal processing.
\end{IEEEbiographynophoto}

\vskip -2\baselineskip plus -1fil
\begin{IEEEbiographynophoto}{Zheng-Hua Tan}
  (M'00--SM'06) received the B.Sc. and M.Sc. degrees in electrical engineering from Hunan University, Changsha, China, in 1990 and 1996, respectively, and the Ph.D. degree in electronic engineering from Shanghai Jiao Tong University (SJTU), Shanghai, China, in 1999.
  He is a Professor in the Department of Electronic Systems and a Co-Head of the Centre for Acoustic Signal Processing Research at Aalborg University, Aalborg, Denmark. He is also a Co-Lead of the Pioneer Centre for AI, Denmark. He was a Visiting Scientist at the Computer Science and Artificial Intelligence Laboratory, MIT, Cambridge, USA, an Associate Professor at the Department of Electronic Engineering, SJTU, Shanghai, China, and a postdoctoral fellow at the AI Laboratory, KAIST, Daejeon, Korea. His research interests include machine learning, deep learning, pattern recognition, speech and speaker recognition, noise-robust speech processing, multimodal signal processing, and social robotics. He has (co)-authored over 200 refereed publications. He is the Chair of the IEEE Signal Processing Society Machine Learning for Signal Processing Technical Committee (MLSP TC). He is an Associate Editor for the IEEE/ACM TRANSACTIONS ON AUDIO, SPEECH AND LANGUAGE PROCESSING. He has served as an Associate/Guest Editor for several other journals. He was the General Chair for IEEE MLSP 2018 and a TPC Co-Chair for IEEE SLT 2016.
\end{IEEEbiographynophoto}

\vskip -2\baselineskip plus -1fil
\begin{IEEEbiographynophoto}{Lars Søndergaard Bertelsen}
  received his M.S.E.E. from Aalborg University in 2000. After that he has worked in the telecommunication industry for several years doing both software, ASIC/FPGA and system design. The recent years he has worked with DSP audio processing in wireless audio products.
\end{IEEEbiographynophoto}

\vskip -2\baselineskip plus -1fil
\begin{IEEEbiographynophoto}{Jens Christian Lindof}
  (M'88) received the M. Sc.EE degree in Telecommunication from Aalborg University, Denmark in 1991. He is currently employed by RTX A/S as its Chief Technology Officer, where he is heading the company’s technology road-mapping activities. Previously among others with Texas Instruments Denmark as Senior Member Technical Staff, site manager and responsible for advanced technology development. Company supervisor for several Industrial PhD students ranging from novel Power Amplifier designs to sound processing algorithms and to the importance of cooperative technical clusters. Co-author of several papers and patents within the area of RF, cellular and audio. Keen on entrepreneurship and thus involved with many start-ups including a handful cofounded over the last 25 years.
  \end{IEEEbiographynophoto}

\vskip -2\baselineskip plus -1fil
  \begin{IEEEbiographynophoto}{Jan Østergaard}
    (S'98--M'99--SM'11) received the M.Sc.E.E.\ degree from Aalborg University, Aalborg, Denmark, in 1999 and the PhD.E.E.\ degree (with cum laude) from Delft University of Technology, Delft, The Netherlands, in 2007. From 1999 to 2002, he worked as an R{\&}D Engineer at ETI A/S, Aalborg, and from 2002 to 2003, he was an R{\&}D Engineer with ETI Inc., VA, USA. Between September 2007 and June 2008, he was a Postdoctoral Researcher at The University of Newcastle, NSW, Australia. He has been a Visiting Researcher at Tel Aviv University, Israel, and at Universidad Técnica Federico Santa María, Valparaíso, Chile. Dr. Østergaard is currently a Full Professor in Information Theory and Signal Processing, Head of the Section on AI and Sound, and Head of the Centre on Acoustic Signal Processing Research (CASPR), at Aalborg University. He has received a Danish Independent Research Council’s Young Researcher’s Award, a Best PhD Thesis award by the European Association for Signal Processing (EURASIP), and fellowships from the Danish Independent Research Council and the Villum Foundations Young Investigator Programme. 
    His research interests are in the areas of acoustic signal processing, statistical signal processing, information theory, joint source-channel coding, and networked control theory.
    He is an Associate Editor for the IEEE TRANSACTIONS ON INFORMATION THEORY
  \end{IEEEbiographynophoto}

\end{document}